\documentclass[12pt]{iopart}
\usepackage{color}
\usepackage{epsfig}
\usepackage{epstopdf}
\usepackage{graphicx}
\usepackage{bm}

\def\micron{\mu \textrm{m}}
\def\diff{\textrm{d}}
\def\wcm2{\ \textrm{W}\,\textrm{cm}^{-2}}

\begin{document}

\title[Micron-scale fast electron filamentation from rear-side optical emission]{Micron-scale Fast Electron Filamentation and Recirculation determined from Rear Side Optical Emission in High Intensity Laser-Solid Interactions}

\author{C Bellei$^1$, S R Nagel$^1$, S Kar$^{2}$, A Henig$^{3,4}$, S Kneip$^{1}$, C Palmer$^{1}$, A S\"{a}vert$^{5}$, L Willingale$^{1,6}$, D Carroll$^{7}$, B Dromey$^2$, J S Green$^{1,8}$, K Markey$^2$, P Simpson$^2$, R J Clarke$^8$, H Lowe$^{8}$, D Neely$^8$, C Spindloe$^{8}$, M Tolley$^{8}$, M Kaluza$^5$, S P D Mangles$^{1}$, P McKenna$^{7}$, P A Norreys$^{1,8}$, J Schreiber$^{1,3,4}$, M Zepf$^2$, J R Davies$^{9}$, K Krushelnick$^{1,6}$, and Z Najmudin$^1$}

\address{$^1$ Blackett Laboratory, Imperial College London, SW7 2AZ, UK}
\address{$^2$ Queen's University of Belfast, Belfast, BT7 1NN, UK}
\address{$^3$ Max-Planck-Institut f\"{u}r Quantenoptik, Garching, Germany}
\address{$^4$ Department f\"{u}r Physik, Ludwig-Maximilians-Universit\"{a}t M\"{u}nchen, Garching, Germany}
\address{$^5$ Institut f\"{u}r Optik und Quantenelektronik, 07743 Jena, Germany}
\address{$^6$ Center for Ultrafast Optical Science (CUOS) University of Michigan, Ann Arbor, Michigan 48109, USA}
\address{$^7$ University of Strathclyde, Glasgow, G4 0NG, UK}
\address{$^8$ Central Laser Facility, STFC Rutherford Appleton Laboratory,
Chilton, Oxon, OX11 0QX,  UK}
\address{$^9$ GoLP, Instituto de Plasmas e Fus\~{a}o Nuclear, Instituto Superior T\'{e}cnico, 1049-001 Lisboa, Portugal}

\ead{claudio.bellei05@imperial.ac.uk}

\begin{abstract}
The transport of relativistic electrons generated in the interaction of petawatt class lasers with solid targets has been studied through measurements of the optical emission from their rear surface. The high degree of polarization of the emission indicates that it is predominantly optical transition radiation. A halo that surrounds  the main region of emission is also polarized, and is attributed to the effect of electron recirculation. The variation of the amplitude of the transition radiation with respect to observation angle provides evidence for the presence of {$\mu$m-size} filaments.
\end{abstract}

\maketitle

\section{Introduction}
The study of electron transport through dense plasmas is an active area of research of importance to many applications.
Relativistic electrons generated in high intensity laser-solid interactions can pass through the critical surface, where the laser energy is mostly absorbed, and continue, free from the influence of the laser, through to the rear surface of the target.
The propagation of this large current of electrons can be affected by self-generated electric and magnetic fields, as well as collisions. A better understanding of this transport is a key issue for the success of the fast ignitor approach to inertial confinement fusion \cite{tabak,atzeni} and may lead to optimization of ion acceleration from laser-irradiated solid targets \cite{clark,snavely}.

Electron transport has been studied
by $K_\alpha$ and XUV emission \cite{stephens,lancaster}, shadowgraphy \cite{tatarakis,borghesi,gremillet1} and optical emission \cite{santos1, baton}.
In particular, the divergence and temporal modulation of the electrons have been investigated by spatially and spectrally resolving rear surface optical radiation  \cite{santos1, baton, jung, popescu, santos2}.
This radiation can be attributed to thermal, synchrotron, transition radiation or {coherent wake emission (CWE) \cite{teubnerprl}}.

These different mechanisms of emission can be differentiated by their polarization characteristics.
Thermal (blackbody) radiation is not expected to be polarized. Synchrotron radiation (SR) is mainly polarized in the plane of motion of the electrons \cite{jackson}.
In our case it would be produced by electrons being pulled back by the electrostatic field that builds up at the back of the target \cite{santos1},
so that the polarization would vary across the emission region, depending on the direction in which electrons travel before restriking the surface.
{CWE should be polarized in the plane of polarization of the laser field and is mainly emitted in the direction of propagation of the laser field \cite{teubnerprl}.} Transition radiation (TR) is also polarized; its properties will be discussed in section \ref{sec:polarization}.

In this paper we present the first {spatially resolved} measurements of the polarization of optical emission from the rear of solid targets irradiated by high intensity lasers.
The radiation is found to be uniformly polarized over the emission region, {with a degree of polarization dependent on the orientation of the target rear surface}, demonstrating that it is predominantly TR. The strong variation of the signal with the orientation of the target rear surface (or, in other words, of the angle of observation) reveals that, at high laser intensities, fast electrons propagate in micron-size filaments.
Moreover, imaging the {TR} far from the laser axis gives direct evidence for the presence of recirculating currents.

This paper is organized as follows. The polarization and general properties of transition radiation are briefly discussed in section 2. In sections 3 we describe the results of two different experimental campaigns, which have been performed on the {\sc{Vulcan}} CPA laser system at the Central Laser Facility. The results of section 3 are further discussed in section 4 and a brief summary of the paper is given in section 5.

\section{Polarization properties of transition radiation}\label{sec:polarization}
 The total emitted energy of the transition radiation per unit angular frequency and unit solid angle can be written as \cite{schroeder}
\begin{eqnarray}
\lefteqn{\frac{\partial^2 W}{\partial\omega\partial\Omega}=\frac{\textrm{e}^2 N}{\pi^2c}
\left[\int \diff^3\textbf{p}\left( \mathcal{E}_\parallel^2 +\mathcal{E}_\perp^2 \right) \right. +}
\nonumber
\\
& &  \ \ \ \ \ \ \ \ \ \  \left. \left(N-1\right) \! \left(\left| \int \!\! \diff^3\textbf{p}\,g(\textbf{p})\mathcal{E}_\parallel F\right|^2
\!\!+ \! \left| \int \!\! \diff^3\textbf{p}\,g(\textbf{p})\mathcal{E}_\perp F\right|^2\right) \right]\!\!,
\label{TR}
\end{eqnarray}
\normalsize
where the first integrals (first line) refer to the \emph{incoherent} component of the transition radiation (ITR) and the second integrals (second line) to the \emph{coherent} one (CTR).
Here $g(\textbf{p})$ is the momentum distribution function, $\mathcal{E}_\parallel$ and $\mathcal{E}_\perp$ are the Fourier-transforms of the electric fields in the plane parallel and perpendicular to the radiation plane (defined by the directions of the target normal and the direction of
observation) and $F$ is a coherence function that takes into account the exact time and position at which electrons reach the interface.
\\ \indent
\begin{figure}
\begin{center}
\includegraphics[width=\textwidth]{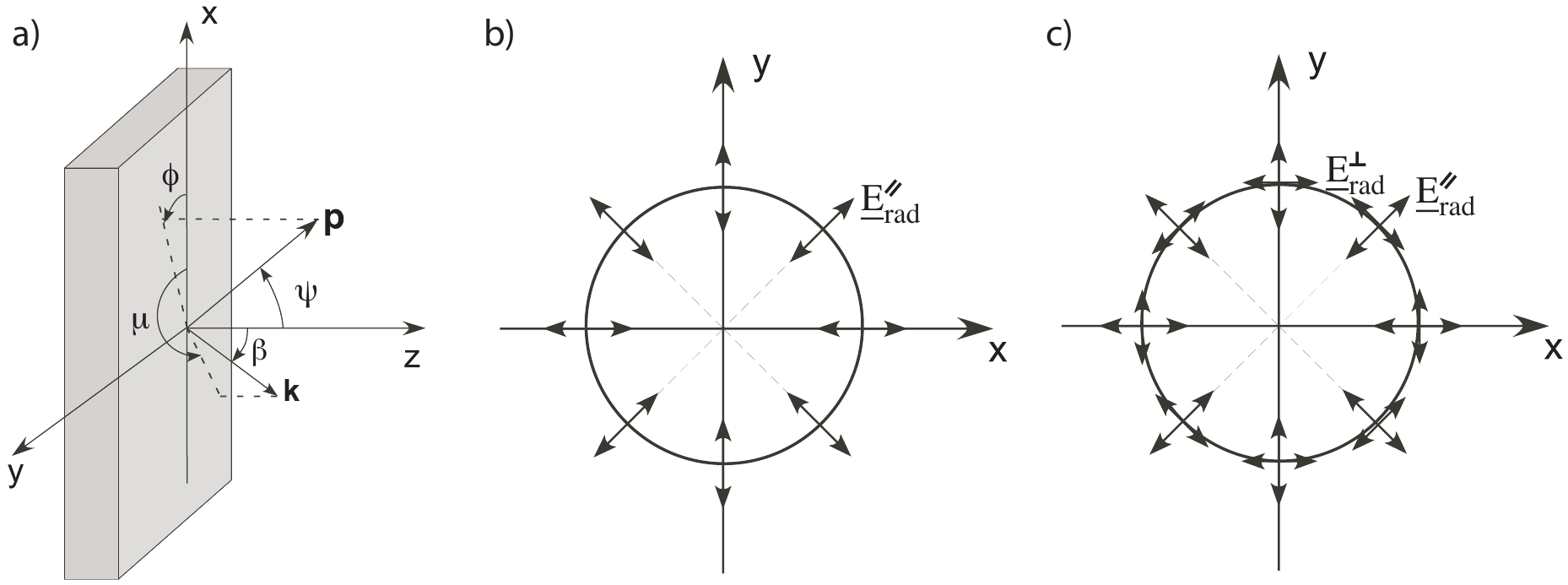}
\end{center}
\caption{a) Geometrical parameters involved in the calculations of TR. Here $\textbf{p}$ is the particle's direction, $\textbf{k}$ is the wave (observation) vector and $z$ is the normal to the target rear surface. b) For a charge
crossing normal to the interface, the radiation is radially polarised.
The magnetic field (not shown) is azimuthal. b) In general,
there is a component of the electric field parallel and normal to
the radiation plane.}
\label{Fig1}
\end{figure}
Expressions for $\mathcal{E}_\parallel$ and $\mathcal{E}_\perp$ can be found in \cite{ter-mikaelian} for the case of two media with generic dielectric constant. However, for our purposes an interface that separates a perfect conductor with the vacuum can be assumed, in which case the Fourier-fields simplify becoming \cite{ter-mikaelian, schroeder}
\begin{eqnarray}
\hspace{-2cm}\mathcal{E}_{\parallel}(\beta, \mu, u, \phi,\psi)= 
\frac{u \cos\psi[u\sin\psi\cos(\phi-\mu)-(1+u^2)^{1/2}\sin\beta]}
{[(1+u^2)^{1/2}-u\sin\psi\cos(\phi-\mu)\sin\beta]^2-u^2\cos^2\psi\cos^2\beta} \ , \label{Epar}
\\ 
\hspace{-2cm} \mathcal{E}_{\perp}(\beta,\mu, u, \phi,\psi)=	
\frac{u^2 \cos\psi\sin\psi\sin(\phi-\mu)\cos\beta}
{[(1+u^2)^{1/2}-u\sin\psi\cos(\phi-\mu)\sin\beta]^2-u^2\cos^2\psi\cos^2\beta} \ , \label{Eperp}
\end{eqnarray}
where $u$ is the normalized momentum, $u=\sqrt{\gamma^2-1}$, and all the other variables are shown in Fig.~\ref{Fig1}a.

In the simplest case of a single particle traversing normal to a plasma-vacuum interface,  transition radiation is $\emph{radially}$
polarised, since $\mathcal{E}_{\perp}=0$ for $\psi=0$. In other words, the electric field oscillates in the radiation plane (Figure
\ref{Fig1}b).
For a particle that does not traverse normal to the interface the radiation is not radially
polarized and a component normal to the radiation plane appears
(Figure \ref{Fig1}c). In this general case, the polarization state is dependent on both the particle's momentum vector $\bf{p}$ and the direction of observation $\bf{k}$.  Thus for a given set of observation angles the polarization state of the radiation carries information on the direction of the particles as they escaped the rear surface.  This will be used for later considerations, to estimate the direction of the electrons from the polarization state of the radiation.

\section{Experimental results}
\begin{figure}[t]
\centerline{\includegraphics[width=\textwidth]{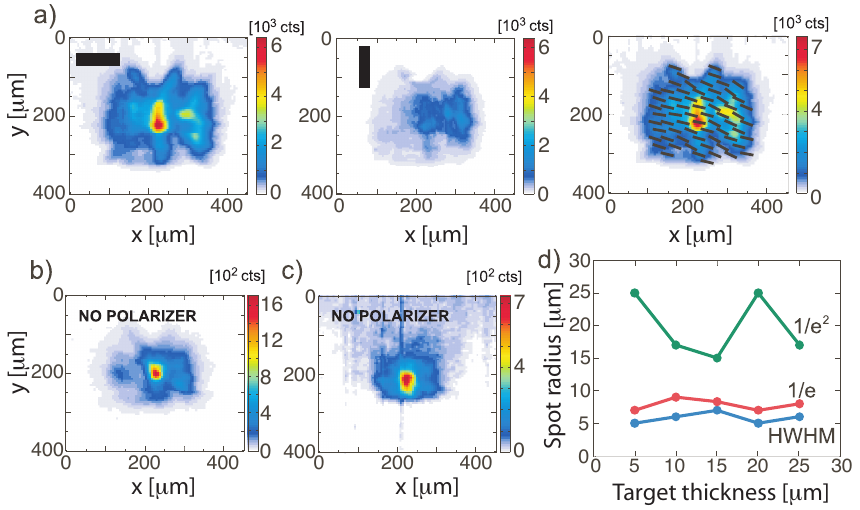}}
\caption{Experiment 1 -  Optical images at  {$2\omega_0$} for (a) $5$, (b) $15$, (c) $25 \ \mu \textrm{m}$ Au targets. (a) Left and central panes correspond to horizontal and vertical polarization respectively, right pane is resulting polarization map. (d) Radius of emission evaluated at $\frac{1}{2}$, $1/e$ and $1/e^2$ of the maximum intensity.}
\label{Fig2}
\end{figure}
The experiments were performed with the {\sc Vulcan} CPA laser operating with wavelength $\lambda_\textrm{0} = 1.054 \, \mu$m and pulse length $580\pm 114\:\textrm{fs}$.
\subsection{Experiment 1}
A series of shots were taken with $\sim 300$ J of energy on target (Experiment 1).
The \textit{p}-polarized laser pulses were focused by a $f/3$ parabola {to a FWHM spot size $w_0 \simeq 8\: \mu\textrm{m}$ and a peak intensity $I \sim 10^{21}\: \textrm{W cm}^{-2}$}  on to a range of flat Au foils of varying thickness at an angle of incidence of $\sim 40^\circ$. The target rear surface was imaged at $\sim 55^\circ$ from target normal and $\sim15^\circ$ from laser axis in the horizontal plane onto two 16 bit CCD cameras equipped with $2\omega_0$ interference filters
and polarizers. The magnification was $\times5$ with a theoretical resolution of $5~\mu\textrm{m}$.
\\
\indent Figure \ref{Fig2}a shows an image of the rear-surface emission from a 5 $\mu \textrm{m}$ Au target. The image is horizontally polarized over the whole region. This is true for the extended halo as well as for the brightest central emission. Hence the emission is not thermal, and the uniformity of the polarization is also inconsistent with it being SR. The measurement \emph{is} however consistent with it being radially polarised TR viewed in the horizontal plane.
That the halo is also TR, indicates that 
the electrons must have been transported transversely by multiple reflections within the target.

If, on the contrary, they had travelled ballistically from the source to an emission point with transverse displacements $> 100\, \mu$m, then they would be travelling at $\sim 90^{\circ}$ to target normal, which would severely inhibit their ability to generate TR. In fact, eqs.~(\ref{Epar}) and (\ref{Eperp}) show that $\mathcal{E}_{\parallel},\ \mathcal{E}_{\perp}\rightarrow0$
for $\psi\rightarrow90^{\circ}$ (electron at grazing incidence). However, the signal at a displacement of $\sim 100\, \mu$m is as much as 20\% of the peak intensity. This implies smaller angles of exit and is thus direct evidence for refluxing of hot electrons which has previously only been deduced indirectly \cite{mackinnon, nilson, mckenna}. Refluxing occurs because of the inhibition of electrons exiting into vacuum by the strong sheath fields that they generate, and is most dominant for thin targets where the surface charge density is greatest. \\
We also note that the TR signal strength decreases with increasing target thickness, as expected, but the spatial extent of the emission remains relatively unchanged (Fig.~\ref{Fig2}d). This is in contrast to previous measurements at lower intensities where increasing emission size with target thickness could be explained by ballistic transport of hot electrons \cite{santos1}.

\subsection{Experiment 2}
A further scan was performed on targets with varying rear-surface wedge angle (Experiment 2). For these shots, an energy of up to $60 \ \textrm{J}$ on target was focused with $\sim 35\%$ of the energy within $w_0 \simeq 6 \ \mu\textrm{m}$, to give $I \sim 8\times 10^{19} \ \textrm{W}\textrm{cm}^{-2}$. In this case the \textit{p}-polarized pulses were incident at a fixed angle of $8^\circ$ to target normal.

The radiation was collected using an $f/3.5$ imaging system, centered at $\sim 33^\circ$ from the normal to the target front (i.e.~$\sim 41^\circ$ to laser axis) in the horizontal plane, with magnification $\times 9.4$ and theoretical resolution $2.8 \ \mu\textrm{m}$. The radiation was split into two orthogonal polarizations with a Wollaston prism and imaged onto the same chip of a $2\omega_0$ filtered $16$ bit CCD camera.

\begin{figure}[h]
\begin{center}
\includegraphics[width=0.8\textwidth]{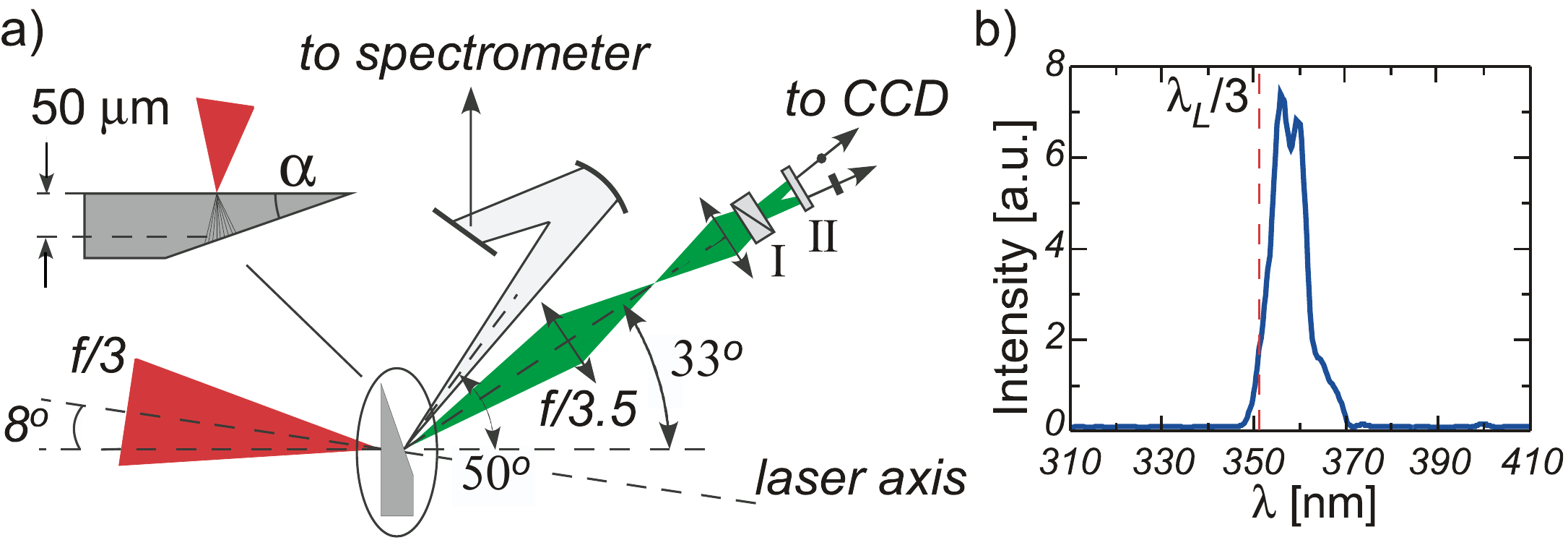}
\end{center}
\caption{Experiment 2 - a) Experimental set-up: {I} Wollaston prism and {II} $2\omega_0$ interference filter; $\alpha$ is the angle of the wedge. b) Spectrum for $\alpha=35^\circ$, showing in detail the 3rd harmonic.}
\label{Fig3}
\end{figure}
The radiation was also spectrally resolved from the NIR to the UV using an optical spectrometer with resolution $\sim 1 \ \textrm{nm}$. Harmonics of the laser frequency $\omega_0$ were recorded in the emission, most strongly at $2\omega_0$, but also at $1\omega_0$, $3\omega_0$ and $4\omega_0$ (e.g.~Fig.~\ref{Fig3}b). This supports the fact that {the} emission is mainly coherent transition radiation (CTR) at the wavelengths of interest, indicating a combination of resonant and $\bf j \times B$ heating \cite{baton}.

Cu wedges with angles of $\alpha = 10^\circ$, 20$^\circ$ and 35$^\circ$ as well as a $50 \, \mu$m flat ($\alpha = 0^\circ$) target were used (Fig.~\ref{Fig3}a). The interaction point was chosen to ensure that the distance from front to rear surface was 50 $\mu\textrm{m}$ for each target, thus keeping the effective foil thickness constant.
The use of wedge targets allowed the angle that the fast electron beam formed to the rear surface to be varied without significantly changing the absorption at the front surface. Hence it can be assumed that the fast electrons produced in the interaction for all these targets had similar properties (temperature, direction, temporal envelope).
As the collection optics were at a fixed position, this also enabled a variation of the angle of observation $\beta$ (shown in Fig.~\ref{Fig1}a) with respect to the normal to the target rear side.

In the following two properties of the data collected in this campaign (Experiment 2) will be discussed, as a function of the angle of observation:
\begin{enumerate}
\item the polarization state of the radiation, from which an estimate for the direction of the electron filaments diagnosed by our imaging system will be inferred;
\item the integrated signal level  recorded on the CCD camera, from which an estimate for the size of the electron beamlets responsible for the CTR signal will be inferred.
\end{enumerate}

\subsubsection{Polarization data - direction of the electron filaments.}
\begin{figure}[h]
\begin{center}
\includegraphics[width=0.9\textwidth]{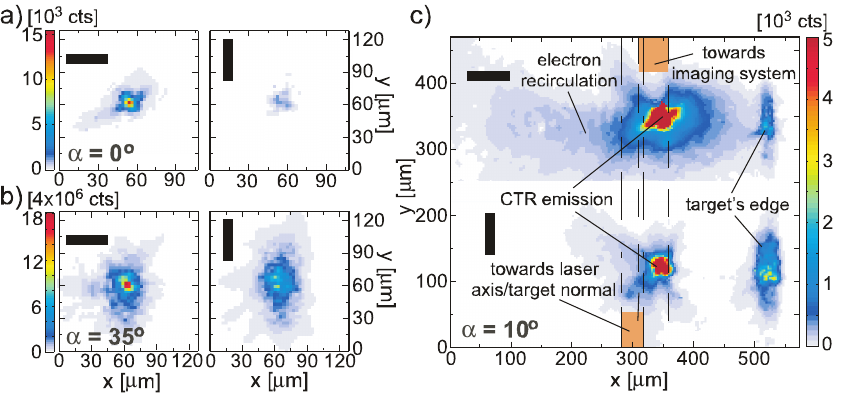}
\end{center}
\caption{Experiment 2 - Polarization analysed images; LHS horizontal polarization, RHS vertical polarization for a) 50 $\mu$m foil ($\alpha=0^\circ$), b) 35$^\circ$ wedge, c) 10$^\circ$ wedge (here top is horizontal and bottom vertical polarization). The center is overexposed to enhance target visibility. }
\label{Fig4}
\end{figure}
The radiation exhibited a high degree of polarization that was dependent on the wedge angle $\alpha$ (Fig.~\ref{Fig4}). As for Experiment 1, these images exhibit a wide polarized halo surrounding the bright central CTR, most clearly seen in Fig.~\ref{Fig4}c. On the right side of this image, for both polarizations, a bright line of emission is seen from the target edge. Both the extended halo and the target-edge emission are at distances far from the initial interaction region. Therefore, these effects can also be attributed to electron recirculation. \emph{Ion emission} from the edge of solid targets has been observed previously \cite{mckenna}, and has also been attributed to the dynamics of electrons transported through the targets by recirculation.

For flat targets the dominant polarization was horizontal, as before.
However, increasing the  wedge angle increased the relative contribution of the vertical polarization as expected for TR. 
A mean direction for the electrons diagnosed by our imaging system can be inferred from the dependence of the polarization state on $\beta$. 

\begin{figure}[t]
\begin{center}
\includegraphics[width=0.4\textwidth]{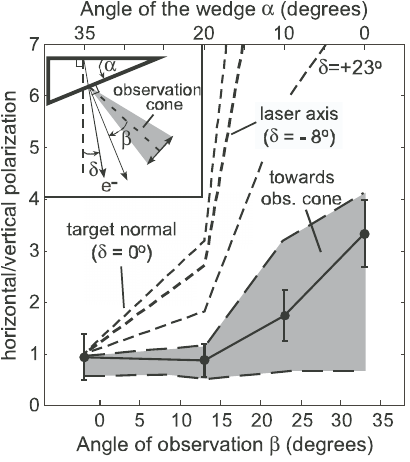}
\end{center}
\caption{Experiment 2 - Variation of polarization ratio with observation angle $\beta$. Experiment (black circles) and theoretical predictions for different directions $\delta$ of electron beams  (dashed lines). The shaded region is obtained assuming electrons directed within the observation cone of the imaging system, $\delta \in [25^\circ, 42^\circ]$.}
\label{Fig5}
\end{figure}

For this analysis, the CTR component of eq.~(\ref{TR}) was integrated over the solid angle of the collection optics. The form factor $F$ was evaluated as shown in the next paragraph \ref{par:beamsize}, however it is not of relevance for the present discussion. Our choice has been to identify the polarization state of the radiation by the ratio of the horizontal to the vertical polarization. For each wedge target, the direction $\delta$ of a collimated beam of electrons is then varied with respect to the \emph{front} side target normal (which was not varied for the different wedge targets). For each of these directions, integration of eq.~(\ref{TR}) over the solid angle of the imaging system gives rise to a particular value of this polarization ratio. This process is then repeated for a different wedge target. Figure \ref{Fig5} shows the results for varying $\delta$ compared to the experimental results at different $\beta$ (or, equivalently, $\alpha$). It should be noted that the dominant polarization is horizontal, since we were observing the radiation mainly along the horizontal plane.

The analysis suggests that the electron beamlets diagnosed by the imaging system were directed within the cone of the collection optics, i.e.~$\delta \in [25^\circ, 42^\circ]$. This is also supported by noting the distance of the main CTR from the laser axis (Fig.~\ref{Fig4}c), which can be determined by measuring the distance
 from the target edge. 
\begin{figure}[b]
\begin{center}
\includegraphics[width=0.8\textwidth]{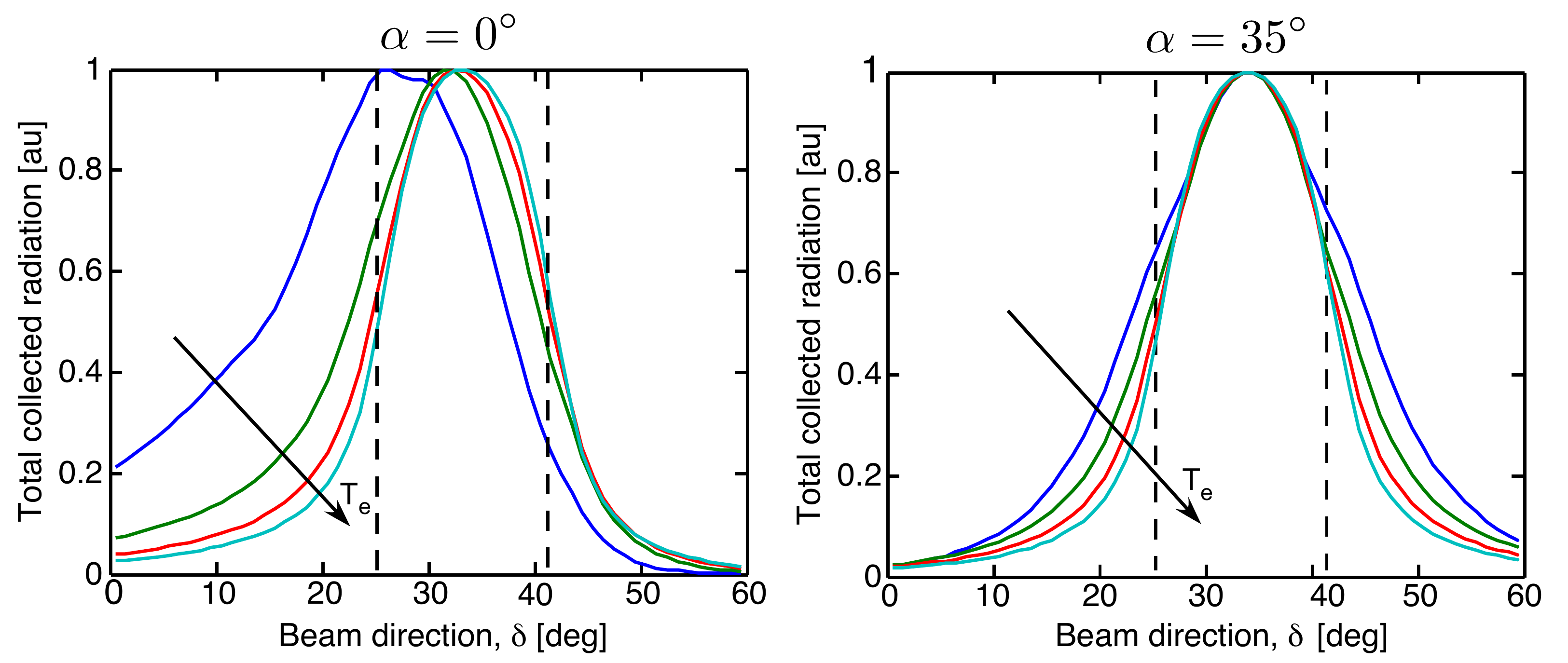}
\end{center}
\caption{Total normalized energy collected by the imaging system as a function
of the electron direction $\delta$ (measured with respect to the \emph{front} side target
normal, as in the inset of Fig.~\ref{Fig5}), for different electron temperatures (blue: 0.5 MeV, green:
2 MeV, red: 4.5 MeV, turquoise: 9.5 MeV). The two broken lines mark
the range of angles included in the cone of the first lens. Each curve
is re-scaled to its maximum value, in general more energy is radiated
at higher temperatures.}
\label{Fig6}
\end{figure}

This conclusion is also supported more directly by noting the distance
of the main CTR from the laser axis in Fig.~ \ref{Fig4}c,
which can be determined by measuring the distance from the target
edge. The position of the signal from the CTR coincides with the position
along the direction of the first lens.

This result is expected since for relativistic
electrons the angular distribution of transition radiation is confined
almost along the particle's direction of propagation. The (theoretical) amount of radiation collected by the imaging
system as a function of the same angle of the electrons $\delta$, and for two
different wedge targets ($\alpha=0^{\circ},35^{\circ}$) is shown in Fig.~\ref{Fig6}. It is clear
that electrons going towards the collection optics (denoted by the
two vertical broken lines) are more easily detected by the imaging
system. These plots are made assuming a Maxwellian distribution of
electrons.

\subsubsection{Signal intensity data - size of the electron filaments.}\label{par:beamsize}
\indent The size of the main bright source of radiation in Fig.~\ref{Fig4} is limited by the resolution of the imaging system. A measure of this size can also be obtained by using the angular distribution of the radiated energy, which for CTR should be strongly dependent on source size.
To interpret the variation of the signal intensity with wedge  angle it is important to retain all the parameters in the CTR component of eq.~(\ref{TR}). In particular, the role of the form factor  $F$ is essential in this case. Physically, it accounts for the phase difference of the waves emitted at the rear of the target.
Therefore it depends on the time and position at which each electron reaches the back surface.  Assuming that a train of impulsive electron bunches is produced at the front surface, that the beam is collimated and that the transverse profile of the electron beam is Gaussian, $F$ is given by
\begin{equation}
	F=G\ \frac{\sin(n_b\,\omega_\textrm{\small{obs}}\,\delta_T/2)}{\sin(\omega_\textrm{\small{obs}}\,\delta_T/2)}
		\exp\left[-{\frac{1}{2}}\left(\frac{\pi D}{\lambda_\textrm{\small{obs}}}\right)^2 \sin^2\beta\right]\ .
	\label{formfactor}
\end{equation}
\begin{figure}[b]
\begin{center}
\includegraphics[width=0.4\textwidth]{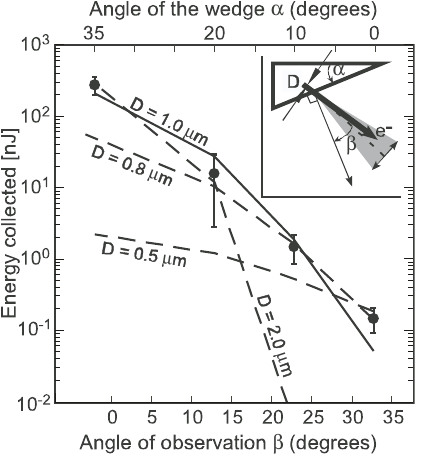}
\end{center}
\caption{Experiment 2 - Variation of signal intensity with $\beta$ (angle between the center of the observation cone and the \emph{rear} side target normal). Experiment (black circles) and theoretical fits (dashed lines, in arbitrary units) for different filament diameters $D$.}
\label{Fig7}
\end{figure}

Here $\delta_T$ is the bunch period, $n_b$ the number of bunches and $D$ the diameter of the beam at $1/\sqrt\textrm{e}$ maximum intensity.
The effect of the quantity $G$ (for which we do not give a full expression since it is not relevant for this discussion) is mainly a decrease in signal strength with target thickness. The second term in eq.~(\ref{formfactor}) gives rise to harmonics of the bunch frequency in the spectrum, as reported previously \cite{baton, popescu, santos2}.
\\
\indent The third term in eq.~(\ref{formfactor}) is the result of the Fourier-transform of the transverse profile of the electron current \cite{schroeder, zheng1}.  It can be understood by analogy with classical diffraction in the far-field (Fraunhofer) approximation.
The radiation emitted from the rear surface can be thought of as the diffraction of a plane wave by a circular aperture with a Gaussian transmission function. When the angle of observation $\beta$ differs from zero, this term leads to a rapid decrease in intensity, depending on the transverse size of the beam $D$. This idea was suggested by Zheng {\em et al}.\ \cite{zheng1} but, to our knowledge, in all previous experiments the observation angle was kept constant. 
\\
\indent The data in Fig.~\ref{Fig7} have been fitted for different beam sizes.
The electron temperature was assumed to be $T_\textrm{e} = 10$ MeV, but the dependence on $T_\textrm{e} $ is weak. {A best fit over all the experimental  data implies FWHM diameter $D = 1 \: \mu$m, however the fall-off in signal strength at large $\beta$ is too strong.} 
A better fit is found if only the data for $\alpha=0^\circ,\;10^\circ,$ and $20^\circ$ are considered. In this case, the best fit is for $D=0.8\ \mu$m. For $D=0.5\ \mu$m the calculated fall-off in signal would be slower than measured.

\section{Discussion}
Numerous numerical results have demonstrated the formation of $\micron$-scale filaments directed over a wide range of angles \cite{pic, evans}.
{For our conditions}, these filaments would have to propagate through $\sim50\ \mu$m of solid Cu, where angular scattering would be expected to cause {rapid transverse expansion.}
Neglecting energy loss and assuming small total angular deflection, this ``beam blooming'', defined as the variance of the transverse beam size due to scattering \cite{li_petrasso}, can be written as
\begin{equation}\label{blooming}
	 B \approx \frac{e^2}{2\sqrt{3\pi}\varepsilon_0}Z\sqrt{n_a\ln\Lambda_s}\frac{s^{3/2}}{pv},
\end{equation}
where $s$ is the distance travelled, $Z$ the atomic number,  $n_a$ the atom number density, $\Lambda_s$ is a term that depends on the electron-atom scattering cross section and $p$, $v$ are the momentum and velocity of the electron, respectively.
This result gives excellent agreement with Monte Carlo modelling \cite{davies}.
Eq.~(\ref{blooming}) indicates that, in order to obtain features {$\simeq1.0\ \mu\textrm{m}$} FWHM ($\equiv 2\sqrt{2\ln 2}B$), electrons with energies {$> 50$ MeV} are required for all realistic values of $\Lambda_s$.
This is well above the ponderomotive potential of the laser ($\sim 4$ MeV) and contradicts the maximum of $30$--$40$ MeV  measured with an electron spectrometer positioned on laser axis in this experiment. \\ \indent
Direct imaging of rear surface emission has recently suggested the presence of filaments as small as $2 \;  \mu$m and with a mean size of $4 \;  \mu$m for intensities of $\sim 10^{19} \ \textrm{Wcm}^{-2}$ onto a range of different targets \cite{storm1,storm2}. However, we show that at higher intensities, the filament size can be below that observable with traditional imaging techniques.
This small scale structure can only be explained by the influence of self-generated electric or magnetic fields inside the target. This could be due to the resistive growth of the magnetic field inside the target \cite{gremillet2}.
Results from the code LSP for parameters close to those of this experiment show no sign of filaments at $0.5$ ps for intensities up to $10^{19}$ Wcm$^{-2}$, but exhibit filaments $1$--$2$ $\mu$m wide at $10^{20}$ W cm$^{-2}$ (see Fig.~12 in \cite{evans}).
In addition, rear surface magnetic focusing \cite{PRE2} and instabilities associated with the electron sheath at the back surface may contribute to the observed small emission size.

\section{Conclusions}
{In conclusion, we have shown that the Polarization of CTR can be used to directly characterise non-linear propagation phenomena of intense electron beams such as recirculation and filamentation with unprecedented spatial resolution. These results imply that the transport is strongly influenced by collective effects at high laser intensity. Understanding the formation and behavior of these small scale structure will be of vital importance for applications of high intensity laser physics.}

\section*{Acknowledgments}
We acknowledge discussions with S. Atzeni and R. J. Kingham and the assistance of the staff of the Central Laser Facility at the Rutherford Appleton Laboratory.


\begin{thebibliography}{}

\end{thebibliography}


\section*{References}
\begin{thebibliography}{100}
\bibliographystyle{unsrt}
\bibitem{tabak} M.~Tabak \textit{et al.}, Phys.~Plasmas \textbf{1}, 1626 (1994)
\bibitem{atzeni} S.~Atzeni, Phys.~Plasmas \textbf{6}, 3316 (1999).
\bibitem{clark} E.~Clark \textit{et al.}, Phys.~Rev.~Lett.~\textbf{84}, 670 (2000).
\bibitem{snavely} R.~A.~Snavely \textit{et al.}, Phys.~Rev.~Lett.~\textbf{85}, 2945 (2000).
\bibitem{stephens} R.~B.~Stephens \textit{et al.}, Phys.~Rev.~E \textbf{69}, 066414 (2004). \bibitem{lancaster} K.~L.~Lancaster \textit{et al.}, Phys.~Rev.~Lett.~\textbf{98}, 125002 (2007).
\bibitem{tatarakis} M.~Tatarakis \textit{et al.}, Phys.~Rev.~Lett.~\textbf{81}, 999 (1998).
\bibitem{borghesi} M.~Borghesi \textit{et al.}, Phys.~Rev.~Lett.~\textbf{83}, 4309 (1999).
\bibitem{gremillet1} L.~Gremillet \textit{et al.}, Phys.~Rev.~Lett.~\textbf{83}, 5015 (1999).
\bibitem{santos1} J.~J.~Santos \textit{et al.}, Phys.~Rev.~Lett.~\textbf{89}, 025001 (2002).
\bibitem{baton} S.~D.~Baton \textit{et al.}, Phys.~Rev.~Lett. \textbf{91}, 105001 (2003).
\bibitem{jung} R.~Jung \textit{et al.}, Phys.~Rev.~Lett.~\textbf{94}, 195001 (2005).
\bibitem{popescu} H.~Popescu \textit{et al.}, Phys.~Plasmas \textbf{12}, 063106 (2005).
\bibitem{santos2} J.~J.~Santos \textit{et al.}, Phys.~Plasmas \textbf{14}, 103107 (2007).
\bibitem{teubnerprl} U.~Teubner \textit{et al.}, Phys.~Rev.~Lett.~\textbf{92}, 185001 (2004).
\bibitem{jackson} Jackson, \textit{Classical Electrodynamics}, John Wiley \& Sons, 1998.
\bibitem{ter-mikaelian} M.~L.~Ter-Mikaelian, \textit{High-Energy Electromagnetic Processes in Condensed Media}, Wiley-Interscience, 1972.
\bibitem{schroeder} C.~B.~Schroeder \textit{et al.}, Phys.~Rev.~E \textbf{69}, 016501 (2004).
\bibitem{mackinnon} A.~J.~Mackinnon \textit{et al.}, Phys.~Rev.~Lett.~\textbf{88}, 215006 (2002).
\bibitem{nilson} P.~M.~Nilson  \textit{et al.}, Phys.~Rev.~E \textbf{79}, 016406 (2009).
\bibitem{mckenna} P.~McKenna \textit{et al.}, Phys.~Rev.~Lett.~\textbf{98}, 145001 (2007).
\bibitem{zheng2} J.~Zheng \textit{et al.}, Phys.~Plasmas \textbf{10}, 2994 (2003).
\bibitem{zheng1} J.~Zheng \textit{et al.}, Phys.~Plasmas \textbf{9}, 3610 (2002).
\bibitem{pic} B. F. Lasinski \textit{et al.}, Phys.~Plasmas \textbf{6}, 2041(1999); H. Ruhl, Plasma Sources Sci. Technol. 11, A154$-$A158 (2002) ; Y. Sentoku \textit{et al.}, Phys.~Rev.~E \textbf{65}, 046408 (2002);  C. Ren \textit{et al.},  Phys.~Rev.~Lett.~\textbf{93}, 185004 (2004); J. C. Adam \textit{et al.}, Phys.~Rev.~Lett.~\textbf{97}, 205006 (2006).
\bibitem{evans} R. G. Evans, High Energy Density Physics, {\bf 2} 35 (2006).
\bibitem{li_petrasso} C.~K.~Li and R.~D.~Petrasso, Phys.~Rev.~E \textbf{73}, 016402 (2006).
\bibitem{davies} J. R. Davies  \textit{et al.}, Phys.~Rev.~E \textbf{56}, 7194, (1997).
\bibitem{storm1} M. Storm  \textit{et al.},  Rev. Sci. Instrum.~\textbf{79}, 10F503 (2008).
\bibitem{storm2} M. Storm  \textit{et al.},  Phys.~Rev.~Lett.~\textbf{102}, 235004 (2009).
\bibitem{gremillet2} L. Gremillet \textit{et al.}, Phys.~Plasmas \textbf{9}, 941 (2002).
\bibitem{PRE2} J. R. Davies, A. R. Bell, and M. Tatarakis, Phys. Rev. E {\bf 59}, 6032 (1999).

\end{thebibliography}
\end{document}